\documentclass[a4paper,11pt]{article}

 \usepackage{amsfonts,amsthm,amssymb}
 \usepackage{latexsym,color, epsf}

 \def\picill#1by#2(#3)
 {\vbox to #2 {\hrule width #1 height 0pt depth 0pt
 \vfill\epsffile{#3}}}

\newcommand{\eq}{\begin{equation}}
\newcommand{\en}{\end{equation}}
\newcommand{\eqa}{\begin{eqnarray}}
\newcommand{\ena}{\end{eqnarray}}

\begin{document}

\setlength{\unitlength}{1mm}

\thispagestyle{empty}


\vspace*{0.1cm}

 \begin{center}

{\bf  Quantum Error Correction Code in the Hamiltonian Formulation}

 \vspace{.2cm}

 Yong Zhang
 \footnote{yong@physics.utah.edu}
 \\[.2cm]

 Department of Physics, University of Utah \\
  115 S, 1400 E, Room 201, Salt Lake City, UT 84112-0830
\\[0.1cm]

\end{center}

\vspace{0.2cm}

\begin{center}
\parbox{14cm}{
\centerline{\small  \bf Abstract}  \noindent\\

The Hamiltonian model of quantum error correction code in the
literature is often constructed with the help of its stabilizer
formalism. But there have been many known examples of nonadditive
codes which are beyond the standard quantum error correction theory
using the stabilizer formalism. In this paper, we suggest the other
type of Hamiltonian formalism for quantum error correction code
without involving the stabilizer formalism, and explain it by
studying the Shor nine-qubit code and its generalization. In this
Hamiltonian formulation, the unitary evolution operator at a
specific time is a unitary basis transformation matrix from the
product basis to the quantum error correction code. This basis
transformation matrix acts as an entangling quantum operator
transforming a separate state to an entangled one, and hence the
entanglement nature of the quantum error correction code can be
explicitly shown up. Furthermore, as it forms a unitary
representation of the Artin braid group, the quantum error
correction code can be described by a braiding operator. Moreover,
as the unitary evolution operator is a solution of the quantum
Yang--Baxter equation, the corresponding Hamiltonian model can be
explained as an integrable model in the Yang--Baxter theory. On the
other hand, we generalize the Shor nine-qubit code and articulate a
topic called quantum error correction codes using
Greenberger-Horne-Zeilinger states to yield new nonadditive codes
and channel-adapted codes.

  }

\end{center}


\begin{tabbing}

\end{tabbing}


\newpage

\section{Introduction}

Quantum information and computation \cite{nc99, mermin07} is a new
interdisciplinary field combining (quantum) physics, (advanced)
mathematics and (modern) computer science, and one of its great aims
is to build a real quantum computer on which quantum algorithms can
successfully run.  Quantum error correction codes are exploited to
protect quantum information from various kinds of noise and perform
a large-scale quantum computation with imperfect quantum gates. They
were originally invented by Shor, Steane, Calderbank, et al.
\cite{shor95, cs96, steane96a, steane96b}, for examples, the Shor
nine-qubit code \cite{shor95}, the Steane seven-qubit code
\cite{steane96a, steane96b}, and the perfect five-qubit code
saturating the quantum Hamming bound \cite{bdsw96, lmpz96}. Additive
quantum codes analogous to classical linear codes can be described
in the stabilizer formalism which have two well known forms
\cite{crss97a,crss97b} and \cite{gottesman97}. Quantum error
correction conditions \cite{bdsw96, kl97} play a fundamental role in
the quantum error correction theory as a good guidance of devising
good quantum error correction codes. Fault-tolerant quantum
computations \cite{shor96, kitaev97, preskill97, preskill98, klz98}
study how to perform reliable quantum computation with the help of
quantum error correction codes, and the threshold theorem
\cite{abo97, aharonov99} can be used to evaluate various proposals
for fault-tolerant quantum computations and guide physicists to
devise reasonable experiments in quantum information and
computation.

Quantum computer is a physical system described by quantum
mechanics, i.e., its dynamics determined by the Hamiltonian in the
Shr{\"o}dinger equation. The focus of the present paper is to study
the Hamiltonian model underlying quantum error correction code. In
Kitaev's toric code \cite{kitaev97}, the Hamiltonian is a linear
combination of elements of the stabilizer group, its degenerate
ground state represents a quantum error correction code, and there
is a gap between its ground state and its first excited state to
protect this code from environment noise. Subsequent to the toric
code is topological quantum computing \cite{kitaev97,fklw03} in
which qubits are anyons (quasiparticles obeying the braid
statistics) and quantum gates form unitary braid representations. In
condensed matter physics, Abelian or non-Abelian anyons can be
created in the fractional quantum Hall effect \cite{sfnss07}.
Recently, encoding quantum information into a subsystem of a
physical system has been found to be a most general method for
protecting quantum information from decoherence, and it is
summarized in operator quantum error correction \cite{kl97,
klv00,klp05}. Similar to the Hamiltonian model of the toric code,
two examples for how to construct a Hamiltonian model for operator
quantum error correction subsystem codes are presented by Bacon
\cite{bacon06}. Remarkably, the Hamiltonian model of the toric code
\cite{dklp02} on a four dimensional lattice and the Hamiltonian
model of the subsystem code \cite{bacon06} on the three-dimensional
cubic lattice may be theoretical candidates for self-correcting
quantum memory where the robust storage of quantum information is
guaranteed by physical properties of the system (i.e., the
Hamiltonian and boundary conditions or topology). Besides,
Hamiltonian models for self-correcting quantum memory,
fault-tolerant quantum computation can be described by a physical
system with a time-dependent Hamiltonian \cite{agp07,nmb07a,nmb07b}.
In this paper, however, we will discuss quantum error correction
code in a different Hamiltonian formalism from what have been
introduced as above. We study quantum states themselves which
represent a quantum error correction code, rewrite them by a unitary
transformation on the product basis, and then derive the Hamiltonian
to determine this unitary transformation. The stabilizer formalism
is not involved in our time-independent Hamiltonian formulation. We
will explain our motivation in detail with simple examples in the
following sections.

In the literature, there have been many new kinds of quantum error
correction approaches, for examples: nonadditive codes
\cite{ssw07,yclo07,cssz07,yco07, gr08} beyond the stabilizer
formalism, entanglement-assisted quantum error correction
\cite{bdh06}, topological color codes \cite{bm07}, subsystem codes
\cite{bc06,ak07}, channel-adapted codes
\cite{lncy97,fletcher07,ls07}, etc. Creating new quantum error
correction code is not a primary goal of this paper, whereas
revisiting quantum error correction codes from a new point of view
is our first point. Quantum error correction code is represented by
a unitary basis transformation matrix from the product basis to the
entangling basis in terms of quantum error correction code, and this
explicitly shows the entanglement nature of quantum error correction
codes (i.e., ``fight entanglement with entanglement" by Preskill
\cite{preskill97,preskill98}). A possible experiment realization of
such a basis transformation matrix can be referred to \cite{klmt99}.
We are only concerned about quantum error correction code itself
instead of its underlying  physical, informational, algebraic,
graphical, topological, and geometrical mechanism. On the other
hand, we will indeed articulate a new topic called ``quantum error
correction codes using Greenberger-Horne-Zeilinger (GHZ) states"
which include as examples the Shor nine-qubit code \cite{shor95},
amplitude damping channel-adapted codes
\cite{lncy97,fletcher07,ls07}, and nonadditive codes
\cite{ssw07,ls07}. It is possible to invent a theoretical framework
for devising quantum error correction codes using GHZ states based
on recognizable properties of GHZ states. GHZ states
\cite{ghz89,ghz90,bpdwz99} are usually assumed to be maximally
entangled states in various entanglement measure theories
\cite{nc99,mermin07}.

 As a remark, our research represents a further development of a
 recent study presented in \cite{aravind97,kl04,zkg05a, zkg05b,
 frw06,zjg06,zg07,zrwwg07, zkw05,zhang06a,zhang06b, zk07}.
 We exploit observations and techniques in these articles
 (especially \cite{zrwwg07}) to study quantum error correction,
 which has not been done before to the best of our knowledge.
 They mainly study interdisciplinary
 themes arising from quantum information and computation, low dimensional
 topology \cite{kauffman02}, the Yang--Baxter equation \cite{yang67,baxter72,
faddeev87} and (almost-complex) differential geometry. For examples:
connections among quantum entanglements, topological entanglements
and geometric entanglements as well as {\em integrable quantum
computation} are discussed in \cite{aravind97,kl04,zkg05a, zkg05b};
a possible link \cite{zrwwg07} between quantum error correction and
topological quantum computing  can be set up via {\em the Bell
matrix}, which forms a unitary braid representation and acts as a
unitary basis transformation matrix from the product basis to GHZ
states. In the sense of quantum information and computation, these
papers articulate {\em integrable quantum computation} as well as
{\em topological-like (partial topological) quantum computation}
\cite{zhang06a,zhang06b,zk07}: the former is a new approach for
quantum computation using {\em integrable models} \cite{zkg05a,
zkg05b,yang67,baxter72, faddeev87} constructed via solutions of the
quantum Yang--Baxter equation (the Yang--Baxter equation with the
spectral parameter); and the latter  exploits low dimensional
topology \cite{kauffman02}, unitary braid gates and non-braid gates.
In our Hamiltonian formulation, Shor's code (and its generalization)
can be explained as a braiding operator, and the corresponding
Hamiltonian model can be recognized as an integrable model.

We hereby summarize our main result which is new to our knowledge.
1) Quantum error correction code is described in a new Hamiltonian
formulation, and the unitary evolution operator at the specific time
is a unitary basis transformation from the product basis to the
quantum error correction code. 2) In our Hamiltonian formulation,
Shor's code and its generalization can be recast as braiding
operators. 3) Quantum error correction codes using GHZ states are
analyzed as an independent topic for yielding new nonadditive codes
or channel-adapted codes. The plan of this paper is organized as
follows. In Section 2, with the helpful formalism of GHZ states,
Shor's nine-qubit code is described in the Hamiltonian formulation
and then explained in the language of the braid group. In Section 3,
we discuss quantum error correction codes using GHZ states, study
generalized Shor's codes as examples, and make comments on
nonadditive codes and channel-adapted codes. Finally, we conclude
this paper by presenting interesting problems for further research.

\section{Notations, motivations and Shor's nine-qubit code}

Notations for most symbols in this paper are introduced, and an
overview is made on physical, informational and mathematical
properties of GHZ states. Hamiltonian formulations of Shor's
nine-qubit code are explored in detail, so that our motivations for
the Hamiltonian formulation of quantum error correction codes
(without exploiting the stabilizer formalism) can be understood
well. Besides, the original motivation takes root in our previous
work \cite{zrwwg07} recasting GHZ states as unitary solutions of the
Yang--Baxter equation (or the braid group relation).

\subsection{Notations}

In quantum information and computation \cite{nc99, mermin07}, a two
dimensional Hilbert space ${\cal H}_2$ over the complex field
${\mathbb C}$ is called a qubit, for example, $\alpha |0\rangle +
\beta |1\rangle$, $\alpha,\beta \in {\mathbb C}$ where $|0\rangle$
and $|1\rangle$ form an orthonormal basis of ${\cal H}_2 \cong
{\mathbb C}^2$ and are usually chosen as eigenvectors of the Pauli
operator $Z$: $Z|0\rangle=|0\rangle$ and $Z|1\rangle=-|1\rangle$.
The Pauli matrices have the conventional form,
\eq X= \left(\begin{array}{cc} 0 & 1 \\
 1 & 0 \end{array}\right), \quad Z=\left(\begin{array}{cc} 1 & 0 \\
 0 & -1 \end{array}\right), \quad Y=ZX=\left(\begin{array}{cc}
 0 & 1 \\ -1 & 0\end{array}
 \right)
\en which denote the bit-flip, phase-flip, and bit-phase flip
operation on a qubit.

The symbol $1\!\! 1_2$ denotes
 a $2$-dimensional identity operator or $2\otimes 2$ identity matrix.
 The notations $A^{\otimes n}$ and $|a\rangle^{\otimes n}$ denote the
 following $n$-fold tensor products,
 \eq
  A^{\otimes n} =\underbrace{A\otimes \cdots A}_n \,, \quad
  |a\rangle^{\otimes n} =\underbrace{|a\rangle \otimes \cdots
  \otimes|a\rangle}_n.
 \en An $n$-fold tensor product, in terms of the identity operator
 $1\!\! 1_2$ and Pauli matrices $X,Y,Z$, has a simpler notation in which
 the lower index of the Pauli matrix labels its position in this tensor
 product, for example, $Z_1 Z_2$ and $X_4 X_5 X_6 X_7 X_8 X_9$ in a
 9-fold tensor product describing the form, respectively,
  \eq
  \label{nota_qec}
Z_1 Z_2 =Z\otimes Z \otimes (1\!\! 1_2)^{\otimes 7}, \quad X_4 X_5
X_6 X_7 X_8 X_9= (1\!\! 1_2)^{\otimes 3} \otimes X^{\otimes 6}.
  \en

The symbols $M$ and $B$ are exploited in the entire paper. The $M$
is an anti-Hermitian operator $M^\dag=-M$ satisfying $M^2=-Id$ with
$Id$ denoting the identity operator, and $B=e^{\frac \pi 4 M}$. But
$M$ and $B$ will have different presentations depending on  where
they appear. If $M$ is a $k$-fold tensor
 product in terms of Pauli matrices $X,Y,Z$, then we introduce
 a notation for an $n$-fold tensor product $M_i$ by
 \eq
 M_i = (1\!\! 1_2)^{\otimes {i-1}} \otimes  M \otimes (1\!\!
  1_2)^{n-k-i+1}, \quad 1 \le i  \le n+1 -k
 \en
where the lower index of $M_i$ can be relabeled for convenience. In
\cite{zg07,zrwwg07}, the symbol $M$ is called the {\em
almost-complex structure} in the (almost-complex) differential
geometry, and it can yield an anti-Hermitian representation of the
{\em extraspecial two-groups}.  As $B=e^{\frac \pi 4 M}$ is a
solution of the Yang--Baxter equation \cite{yang67,baxter72} and can
generate a unitary braid representation $\pi_n$ of the Artin braid
group ${\cal B}_n$ \cite{kauffman02}, it is called the {\em Bell
matrix}, see \cite{zkg05a,zkg05b,zjg06,zg07,zrwwg07}. Since unitary
braids are used as quantum gates in topological quantum computing
\cite{kitaev97,fklw03} and extraspecial two-groups are exploited in
quantum error correction \cite{crss97a,crss97b}, the formula
$B=e^{\frac \pi 4 M}$ suggests there may exist a link between
quantum error correction and topological quantum computing, which is
the main proposal of our previous paper \cite{zrwwg07}.

\subsection{An overview on Greenberger-Horne-Zeilinger states}

The GHZ states \cite{ghz89,ghz90,bpdwz99} are a multipartite
generalization of the bipartite maximally entangled Bell states, and
are defined by various known entanglement measures as maximally
entangled. They play crucial roles in both fundamental problems and
practical applications of quantum information theory. The Bell
theorem \cite{bell64} for incompatibility between quantum theory and
classical deterministic local models is expressed in the form of
inequalities (Bell inequalities) among various statistical
correlations, whereas the GHZ theorem \cite{ghz89,ghz90} asserts
that the quantum correlations represented by the GHZ states allows
us to describe the Bell theorem in terms of equalities and to test
the expected incompatibility only by perfect correlations. The GHZ
states are the simplest multipartite maximally entanglement sources,
and have been widely exploited in the study of quantum information
and computation, for examples, multiparty quantum key distributions
\cite{ekert91} and quantum teleportation \cite{bbcjpw93}. Besides,
GHZ states are often called \textit{cat states} acting as ancillas
in fault-tolerant quantum computation \cite{shor96, preskill97,
preskill98, klz98}.

Besides the fundamental importance of GHZ states in quantum physics
and information, they have been found to posses (beautiful and deep)
topological, algebraic and geometric properties. Aravind
\cite{aravind97} observed that there are similarities between the
GHZ states and knot configurations \cite{kauffman02}, by identifying
the measurement of a specific state of a particle with cutting the
corresponding link component. Furthermore,
 we describe GHZ states by the Bell matrix
$B=e^{\frac \pi 4 M}$ which has become a common topic among quantum
error correction, topological quantum computing, the braid group,
the Yang--Baxter equation, and (almost-complex) differential
geometry \cite{kl04,zkg05a, zkg05b, frw06,zjg06,zg07,zrwwg07}. The
GHZ states of $n$-qubit are defined to have the form, \eq
\label{ghz}
  \frac 1 {\sqrt 2} (|s\rangle \pm |\bar{s}\rangle), \quad
  s=i_1i_2\cdots i_n, \quad \bar{s}=\bar{i}_1 \bar{i}_2\cdots \bar{i}_n
\en where $i_j=0,1$, $j=1,\cdots, n$ and $i_j + \bar{i}_j=1$ with
the Abelian addition modulo 2. In \cite{zrwwg07}, the GHZ states of
the $n$-qubit are generated by the exponential function $B=e^{\frac
\pi 4 M}$ acting on the product basis, for example,
 \eq
\frac 1 {\sqrt{2}} (|0\rangle^{\otimes n} + |1\rangle^{\otimes n})
 = B|0\rangle^{\otimes n},  \quad \frac 1 {\sqrt{2}} (|0\rangle^{\otimes n}
 -|1\rangle^{\otimes n})= B^{-1} |0\rangle^{\otimes n},
 \en
where $M=-Y \otimes X^{\otimes n}$ and  $B^{-1}=e^{-\frac \pi 4 M}$
denotes the inverse of the $B$ matrix. Note that $B=e^{\frac \pi 4
M}$ is called {\em the Bell matrix} only if it is a unitary solution
of the Yang--Baxter equation \cite{yang67,baxter72} (or forms a
unitary braid representation \cite{kauffman02}).

\subsection{Hamiltonian formulation of Shor's nine-qubit code}

The first quantum error correction code was found by Shor
\cite{shor95}, and it describes a logical qubit state in terms of
nine-qubit states in the way \eqa && |0\rangle \to |0\rangle_L
=\frac {1} {2\sqrt{2}} (|000\rangle+|111\rangle) \otimes
(|000\rangle+|111\rangle) \otimes (|000\rangle+|111\rangle),
\nonumber\\
 &&  |1\rangle \to |1\rangle_L =\frac {1}
{2\sqrt{2}} (|000\rangle - |111\rangle) \otimes (|000\rangle
 -|111\rangle) \otimes (|000\rangle -|111\rangle)
\ena where two GHZ states $\frac 1 {\sqrt 2} (|000\rangle \pm
|111\rangle)$ are repeatedly used and hence Shor's code is also
called the repetition code. In the stabilizer formalism
\cite{gottesman97} for additive quantum error correction codes, the
encoded logical qubit $\alpha |0\rangle_L + \beta |1\rangle_L$,
$\alpha, \beta \in {\mathbb{C}}$ is uniquely determined as  a common
eigenvector with the eigenvalue $1$ of the following eight
stabilizer operators,
  \eq
 Z_1 Z_2, Z_2 Z_3, Z_4 Z_5, Z_5 Z_6, Z_7Z_8, Z_8 Z_9, X_1 X_2 X_3 X_4
 X_5 X_6, X_4 X_5 X_6 X_7 X_8 X_9
  \en
  which form an Abelian group called the stabilizer group. The
  logical bit-flip operation $\overline{X}$ and logical phase-flip
  operation $\overline{Z}$ usually have the form \eq \label{shor_log}
 \overline{X} =Z^{\otimes 9}, \quad \overline{Z} =X^{\otimes 9}.
\en

In the following, we firstly describe Shor's nine-qubit code in our
Hamiltonian formulation and then present an interpretation in the
language of the braid group. Denote the symbol $M$ by $M=-Y\otimes
Y\otimes Y$ in this subsection. It is a transition operator between
$|000\rangle$ and $|111\rangle$,
 \eq
 M |000\rangle = |111\rangle, \quad M |111\rangle =-|000\rangle
 \en
due to $-Y|0\rangle=|1\rangle$ and $-Y|1\rangle=-|0\rangle$, and it
is an anti-Hermitian operator $M^\dag=-M$ determining $e^{\frac \pi
4 M}$ to be a unitary operator,
 \eq
  e^{\frac \pi 4 M}=\frac 1 {\sqrt 2} (1\!\! 1_2^{\otimes 3} + M),
   \quad M^2=-1\!\! 1_2^{\otimes 3}.
 \en
Furthermore,  introduce $M_1, M_4, M_7$ as \eqa M_1= M \otimes 1\!\!
1_2^{\otimes 6}, \quad \quad M_4 =1\!\! 1_2^{\otimes 3} \otimes M
\otimes 1\!\! 1_2^{\otimes 3}, \quad M_7 =1\!\! 1_2^{\otimes 6}
\otimes M \ena and denote the summation of $M_1, M_4, M_7$ by
$M_t=M_1+M_4 + M_7$. After some simple algebra, the Shor nine-qubit
code has a new compact formulation,
 \eq
 |0\rangle_L =e^{\frac \pi 4 M_t} |0\rangle^{\otimes 9}, \quad
 |1\rangle_L =e^{-\frac \pi 4 M_t} |0\rangle^{\otimes 9}
 \en
which leads to the bit-phase flip operation $\overline{Y}$ on the
 encoded qubit,
 \eq
  \label{ham_shor}
 |0\rangle_L = e^{\frac \pi 2 M_t} |1\rangle_L, \quad
  |1\rangle_L = e^{-\frac \pi 2 M_t} |0\rangle_L.
 \en
Namely, $\overline{Y}$ has the form \eq
 \overline{Y} =e^{\frac \pi 2
  M_t} =M_1 M_4 M_7 =-Y^{\otimes 9}
\en where the minus sign in the front of $Y^{\otimes 9}$ partly
explains the reason why in the literature the logical bit-flip
operator $\overline{X}$ and the logical phase-flip operator
$\overline{Z}$ have the formulation (\ref{shor_log}) because of
$\overline{Y} =\overline{Z}\, \overline{X}$ and $XZ=-ZX$.

Now let us associate the formulation (\ref{ham_shor}) of the Shor
nine-qubit code with the Hamiltonian defined by $H=i M_t$. A logical
qubit basis, $|0\rangle_L$ and $|1\rangle_L$, is explained as a
unitary evolution of the product basis state $|0\rangle^{\otimes 9}$
(or $|1\rangle^{\otimes 9}$), with the unitary evolution operator
$U(\theta) =e^{-i\theta H}$ determined by the Hamiltonian $H$,
namely,
 \eq
 |0\rangle_L = U(\frac \pi 4)
     |0\rangle^{\otimes 9}, \quad
  |1\rangle_L = U(-\frac \pi 4) |0\rangle^{\otimes 9}
   = -U(\frac \pi 4) |1\rangle^{\otimes 9}
 \en
which leads that the encoded qubit $\alpha |0\rangle_L + \beta
|1\rangle_L$ has a form
 \eq
 \alpha |0\rangle_L + \beta
|1\rangle_L =U(\frac \pi 4) (\alpha |0\rangle^{\otimes 9} - \beta
|1\rangle^{\otimes 9}), \quad \alpha, \beta\in {\mathbb C}.
 \en
The Shr{\"o}dinger equation has the form with the unitary
evolutional solution $U(\theta)$,
 \eq
  \label{sch}
 i \frac {\partial} {\partial \theta} \psi(\theta) =H \psi(\theta),
 \quad H=-i (Y_1 Y_2 Y_3+ Y_4 Y_5 Y_6 + Y_7 Y_8 Y_9 )
 \en
where the Planck constant $\hbar=1$, $\theta$ is regarded as the
time variable and $\psi(\theta)$ represents the wave function in
quantum mechanics. The three-body spatially local Hamiltonian $H$
can be simulated on various lattice models, but all of which may be
equivalent to a one-dimensional spin chain because only the $Y$
operation is involved in this Hamiltonian. Note that we do not
impose any boundary conditions on the Shr{\"o}dinger equation
(\ref{sch}), mainly because they are determined by which type of
lattice model to be chosen.

In our Hamiltonian formulation, obviously, Shor's nine-qubit code is
not an eigenvector of the Hamiltonian $H$ and is not its ground
state. This is essentially different from Hamiltonian models
\cite{kitaev97}\cite{bacon06} respectively for the toric code and
subsystem code, which are constructed with the help of the
stabilizer formalism. In spite of this fact, we expect our
Hamiltonian formulation to be helpful for exploring the topics of
how to encode and decode Shor's code, how to perform the
fault-tolerant quantum computation with Shor's code, and especially
how to construct Hamiltonian model of nonadditive code without using
the standard stabilizer formalism. In the next subsection,
furthermore, we will recast the Shor nine-qubit code as a braiding
operator,  and explain the motivation for our Hamiltonian
formulation from the different point of view.

\subsection{Artin braid group, Yang--Baxter equation
and Shor's code}

The Artin braid group ${\cal B}_n$  \cite{kauffman02} has the
generators $\sigma_i$, $i=1,\cdots, n-1$ satisfying the braid group
relations 1) and commutative relations 2):
 \eq
 1)\,\, \sigma_i \sigma_{i+1} \sigma_i = \sigma_{i+1} \sigma_i \sigma_i,
 \quad 2)\,\, \sigma_{i} \sigma_j =\sigma_j \sigma_i, \quad |i-j|>1.
 \en
In terms of the identity matrix $1\!\! 1_2$ and a $k$-fold tensor
product $M$ involving Pauli matrices, a unitary braid representation
$\pi_n$ of the Artin braid group ${\cal B}_n$ can be constructed in
the way,  \eq
 B_i \equiv\pi_n(\sigma_i) =
 (1\!\! 1_2)^{\otimes i-1} \otimes e^{\frac \pi 4 M} \otimes
 (1\!\! 1_2)^{\otimes n-k-i+1}, \quad 1 \le i  \le n+1 -k.
\en where $e^{\frac \pi 4 M}$ satisfies the Yang--Baxter equation
\cite{yang67,baxter72,zrwwg07}, \eq \label{gybe}
 (G \otimes Id )(Id \otimes G)
 (G \otimes Id)=(Id \otimes G)
 (G \otimes Id )(Id \otimes G), \en
where $G=e^{\frac \pi 4 M}$ and $Id=1\!\! 1_2$ and which is a
presentation of the braid group relation $\sigma_i \sigma_{i+1}
\sigma_i = \sigma_{i+1} \sigma_i \sigma_i$, see \cite{zrwwg07} for
more details. Besides, the quantum Yang--Baxter equation is defined
as the Yang--Baxter equation with the spectral parameter,
 \eq
 \label{gqybe}
 (G(x) \otimes Id )(Id \otimes G(xy))
 (G(y) \otimes Id)=(Id \otimes G(y))
 (G(xy) \otimes Id )(Id \otimes G(x)),
 \en
where $x, y$ are the spectral parameter and which can be referred to
our previous papers \cite{zkg05a, zkg05b, zjg06,zg07,zrwwg07}. It is
well known that in the literature an {\em integrable model} can be
constructed using a solution of the quantum Yang--Baxter equation,
see \cite{faddeev87}.

In Figure 1, the diagram without boxes around crossings is a
diagrammatic representation of the  Yang--Baxter equation
(\ref{gybe}) or (\ref{gqybe}) in which  different strands are
allowed to represent Hilbert spaces of different dimensions, for
example, thin strands acting on ${\cal H}_2$ (i.e., a qubit) and
thick strands acting on the tensor product of ${\cal H}_2$ (i.e., at
least two qubits). On the other hand, $G=e^{\frac \pi 4 M}$ can be
explained as a universal quantum gate \cite{ bb02, kl04} which
transforms a separate state to a maximally entangled state, for
example, a GHZ state. The diagram with boxes is hence regarded as a
diagrammatical identity between two quantum circuits where universal
quantum gates are labeled by boxes with inputs and outputs. In other
words, we are presenting two sorts of interpretations on the same
diagrammatical object: the one is the Yang--Baxter equation (or the
braid group relation) \cite{kauffman02,yang67,baxter72, faddeev87}
and the other is quantum information and computation
\cite{nc99,mermin07}. We are trying to build a bridge between
quantum information theory and Yang--Baxter equation (or low
dimensional topology), which is the underlying motivation of a
series of papers \cite{zkg05a, zkg05b, zjg06,zg07,zrwwg07,
zkw05,zhang06a,zhang06b, zk07}. For examples, we have recognized
{\em Werner states} \cite{werner89} as a sort of rational solutions
of the quantum Yang--Baxter equation \cite{zkw05}; in the
formulation of GHZ states \cite{zg07,zrwwg07}, found mathematical
structures including unitary braid representations, extraspecial
two-groups and almost-complex structure; described quantum
teleportantion protocol \cite{bbcjpw93} by a braiding operator
\cite{zhang06a,zhang06b}; and recast diagrammatical quantum
information protocols involving maximally bipartite entangled states
as the extended Temperley--Lieb diagrammatical configuration
\cite{zhang06a,zhang06b,zk07}. In the following, we refine the braid
group relation from our Hamiltonian formulation of the Shor
nine-qubit code.

\begin{figure}
\begin{center}
\epsfxsize=12.cm \epsffile{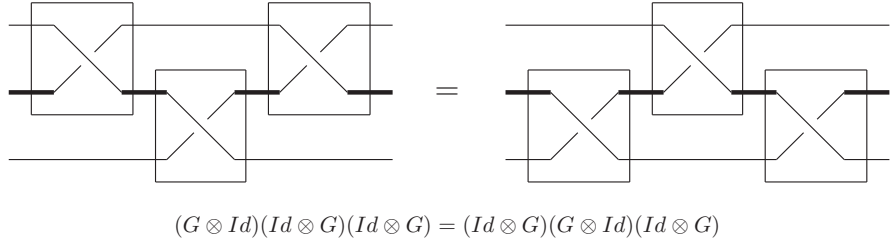} \caption{Yang--Baxter equation
and quantum circuits} \label{fig1}
\end{center}
\end{figure}

Choose $M$ to be the form $M=-Y\otimes X\otimes X$ and construct the
unitary braid representations $B_1, B_4, B_7$ by \eqa B_1 =e^{\frac
\pi 4 M} \otimes 1\!\! 1_2^{\otimes 6}, \quad B_4 =1\!\!
1_2^{\otimes 3} \otimes e^{\frac \pi 4 M} \otimes 1\!\! 1_2^{\otimes
3}, \quad B_7 =1\!\! 1_2^{\otimes 6} \otimes e^{\frac \pi 4 M}, \ena
and then rewrite Shor's nine-qubit code into a compact formulation
\eq |0\rangle_L =B_1 B_4 B_7 |0\rangle^{\otimes 9}, \quad
 |1\rangle_L =B^{-1}_1 B^{-1}_4 B^{-1}_7 |0\rangle^{\otimes 9}.
\en The $B_1, B_4, B_7$ are presentations of the braid group
generators, and have a diagrammatic representation, see Figure 2.
Every straight strand denotes a two-dimensional Hilbert space ${\cal
H}_2$ (i.e., a qubit), and the crossing denotes $e^{\frac \pi 4 M}$
and acts on ${\cal H}^{\otimes 3}_2$. For convenience, we can assign
${\cal H}_2$ to the first strand of the crossing and ${\cal
H}^{\otimes 2}_2$ to the second strand. Therefore, we read the Shor
nine-qubit code from the braiding configuration as a product of
three braids $B_1, B_4, B_7$.

\begin{figure}
\begin{center}
\epsfxsize=12.cm \epsffile{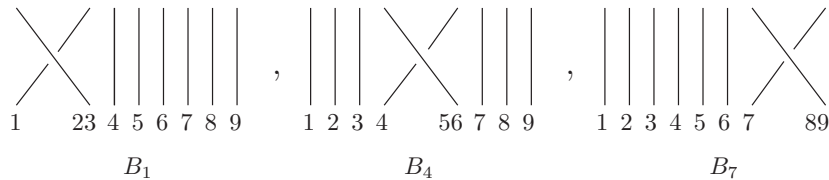} \caption{Artin braid
group generators for Shor's nine-qubit code} \label{fig2}
\end{center}
\end{figure}

On the other hand, in terms of $M_t=M_1 + M_4 + M_7$ where $M_1,
M_4, M_7$ have a form using the convention (\ref{nota_qec}),
 \eq
M_1=-Y_1X_2 X_3, \quad M_4=-Y_4 X_5 X_6, \quad M_7=-Y_7 X_8 X_9,
 \en
we obtain the other Hamiltonian formulation of Shor's code similar
to (\ref{ham_shor}) but the logical bit-phase flip operation
 $\overline{Y}$ given by
 \eq
 \overline{Y} = -Y_1X_2 X_3 Y_4 X_5 X_6 Y_7 X_8 X_9.
 \en
In view of our previous work on {\em integrable quantum computation}
\cite{zkg05a, zkg05b, zjg06,zg07,zrwwg07}, the unitary evolution
operator $U(\theta)=e^{-i\theta H}$ with $H=i M_t$ is a solution of
the quantum Yang--Baxter equation (\ref{gqybe}) with the spectral
parameter $\theta$, and hence the corresponding Hamiltonian model is
an integrable model determined by the solution of the quantum
Yang--Baxter equation \cite{yang67,baxter72,faddeev87}.

 As a remark about our unitary braiding description of Shor's
 nine-qubit code, it provides an interesting example for the
 observation \cite{zrwwg07} that there exists a possible link between
 quantum error correction and topological quantum
 computing via the unitary braid representation using anti-Hermitian
 representation of extraspecial two-groups. Besides, it is possible
 to make connections between our work and a geometric paradigm of
 Shor's code \cite{fm98}.

 \section{Quantum error correction codes using GHZ states}

{\em Quantum error correction codes using GHZ states} are defined as
those codes in terms of GHZ states or tensor products of GHZ states,
while {\em generalized Shor's codes using GHZ states} are specified
to be {\em repetition codes} as a straightforward generalization of
the Shor nine-qubit code, i.e., only involving GHZ states of the
type,
 $1/ {\sqrt 2} (|0\rangle^{\otimes n} \pm|1\rangle^{\otimes n})$.
They are degenerate quantum error correction codes  in which
different errors may be corrected by the same correction operation.
Besides the Shor nine-qubit code, some nonadditive codes
\cite{ssw07, ls07} and channel-adapted codes
\cite{lncy97,fletcher07, ls07} are good examples for quantum error
correction codes using GHZ states. The motivation for our
articulation of this topic is based on the observation that GHZ
states have nontrivial physical, informational, and mathematical
properties (see our overview on GHZ states in Subsection 2.2), and
these properties are expected to be very helpful in the construction
of interesting quantum error correction codes using GHZ states.

In the literature, the triple $[n,k,d]$ of natural numbers
represents a quantum error correction code \cite{cs96}. It is a
unitary mapping $\cal Q$ from ${\cal H}_2^{\otimes k}$ to ${\cal
H}_2^{\otimes n}$ ($k\le n$), and encodes information of $k$-qubit
quantum states into $n$-qubit quantum states in order to detect and
correct $t$-qubit errors, where $t=\frac {d-2} 2$ for $d$ even and
$t=\frac {d-1} 2$ for $d$ odd. Quantum states in this
$2^k$-dimensional subspace ${\cal Q} {\cal H}_2^{\otimes k}$ are
called {\em codewords}. For example, Shor's nine-qubit code is
usually denoted by $[9,1,3]$ and it can correct any single-qubit
errors, while the logical qubit $\alpha |0\rangle_L + \beta
|1\rangle_L$, $\alpha, \beta \in {\mathbb C}$ represents a codeword
in Shor's code.

In this section, for simplicity, we denote quantum error correction
codes $[n,k,d]$ by $[n,k]$, and focus on whether or not they are
able to correct arbitrary single-qubit errors, i.e., $t=1$. For
example, the Shor code is denoted by $[9,1]$. Note that we emphasize
our perspectives of constructing {\em quantum error correction codes
using GHZ states} and describe them in our Hamiltonian formulation,
but leave the systematic construction of procedures for encoding,
error syndrome measurement, error correction and decoding of these
codes elsewhere.

 \subsection{Generalized Shor's codes using GHZ states}

First of all, we introduce symbols and names necessary for
constructing generalized Shor's codes using GHZ states. The {\em
type} of a GHZ state $\frac 1 {\sqrt 2}(|s\rangle \pm
|\bar{s}\rangle)$ (\ref{ghz}) is specified by the number of its
qubits and the sign $\pm$. Generalized Shor's codes exploit
different types of GHZ states to encode one qubit into codewords of
$N$-qubit. The same GHZ states form a {\em block} in the codeword,
and different blocks correspond to orthogonal subspaces of ${\cal
H}_2^{\otimes N}$. The symbol $I$ is the number counting types of
GHZ states (or the number of blocks) in the codeword; $i$ denotes
the $i$-th type of GHZ state (or $i$-th block) in the codeword,
$i=1,\cdots, I$; $q_i$ counts the number of qubits in the $i$-th
type of GHZ state; $n_i$ is the repetition number of the $i$-th type
of GHZ state in the $i$-th block of the codeword; $B=\sum_{i=1}^I
n_i$ counts the number of GHZ states in a codeword; $N=\sum_{i=1}^I
n_i q_i$ is the total number of qubits in the codeword.

For example, the Shor nine-qubit code exploits  GHZ states of
 three-qubit and repeatedly use them three times in a codeword: $I=1$,
$q_1=3$, $n_1=3$, $B=3$ and $N=9$. On the other hand, $B=3$ is the
number counting all phase-flip single-qubit errors (i.e.,
$Z$-errors), and $2N=18$ counts all bit-flip and phase-flip errors
(i.e., $X$-errors and $Y$-errors). Hence Shor's code satisfies the
quantum Hamming bound condition,
  \eq
   \label{ham_bound}
  2^t (1+ B + 2 N ) \le 2^N
   \en
where $t=1$, and which suggests the Shor code as a degenerate code
and able to correct some two-qubit errors. Now let us discuss an
application of the above bound inequality at $t=1$. Obviously,  the
lowest bound for $N$ has to be $N\ge 5$.

At $N=5$, we can either construct a logical qubit using GHZ states
of five-qubit,
 \eq
 |0\rangle_L = \frac 1 {\sqrt 2} (|0\rangle^{\otimes 5} +
  |1\rangle^{\otimes 5}), \quad |1\rangle_L =\frac 1 {\sqrt 2}
  (|0\rangle^{\otimes 5} - |1\rangle^{\otimes 5})
 \en
where $B=1$, $n_1=1$ and $q_1=5$, or have a logical qubit in terms
of Bell states (i.e., GHZ states of two-qubit) and GHZ states of
three-qubit,
 \eq
 |0\rangle_L = \frac 1 2  (|0\rangle^{\otimes 2} +
  |1\rangle^{\otimes 2})  (|0\rangle^{\otimes 3} +
  |1\rangle^{\otimes 3}), \quad |1\rangle_L = \frac 1 2
  (|0\rangle^{\otimes 2} -|1\rangle^{\otimes 2})
  (|0\rangle^{\otimes 3} -|1\rangle^{\otimes 3})
 \en
where $B=2$, $n_1=n_2=1$, $q_1=1$ and $q_2=1$.

At $N=6$, we can construct a logical qubit in terms of GHZ states of
six-qubit (or four-qubit or three-qubit) or Bell states, for
examples,
 \eqa
 && |0\rangle_L = \frac 1 {2\sqrt 2} (|0\rangle^{\otimes 2} +
  |1\rangle^{\otimes 2})(|0\rangle^{\otimes 2} +
  |1\rangle^{\otimes 2})(|0\rangle^{\otimes 2} +
 |1\rangle^{\otimes 2}), \nonumber\\
 && |1\rangle_L =\frac 1 {2\sqrt 2}
  (|0\rangle^{\otimes 2} -|1\rangle^{\otimes 2})
  (|0\rangle^{\otimes 2} - |1\rangle^{\otimes 2})
  (|0\rangle^{\otimes 2} -|1\rangle^{\otimes 2}).
 \ena
where $B=3$, $n_1=3$ and $q_1=3$.

In view of our Hamiltonian formulation of the Shor nine-qubit code,
generalized Shor's codes have the following Hamiltonian formulation,
\eq |0\rangle_L =e^{\frac \pi 4 \sum_{i=1}^I M_t^{(i)}}
|0\rangle^{\otimes N}, \quad |1\rangle_L = e^{-\frac \pi 4
\sum_{i=1}^I M_t^{(i)}} |0\rangle^{\otimes N} \en where the
Hamiltonian  $H^{(i)}=\sqrt{-1}M_t^{(i)}$ determines how to yield
the $i$-th type of GHZ state in the $i$-th block of the codeword and
has the form
 \eq
 H^{(i)} =\sqrt{-1} \sum_{j=1}^{n_i} M_j^{(i)}.
 \en
Different $M_j^{(i)}$ are commutative with each other because the
 corresponding subspace of each type of GHZ state in the codeword is
 orthogonal with those for other types. Note that the lower indices of
$M_j^{(i)}$ are different from that exploited in the Hamiltonian
formations of the Shor nine-qubit code in Subsections 2.3 and 2.4.

Note on a link between the cat code  $[n,1]$ and quantum error
correction codes using GHZ states. The cat code $[n,1]$ has the form
 $|0\rangle_L=|0\rangle^{\otimes  n}$ and $|1\rangle_L
=|1\rangle^{\otimes n}$ leading to
 \eq
 \frac 1 {\sqrt 2} (|0\rangle_L \pm |1\rangle_L )
  =e^{\pm\frac \pi 4 M} |0\rangle_L, \quad |0\rangle_L + |1\rangle_L
   =M (|0\rangle_L -|1\rangle_L)
 \en
in which $M$ can take a form $M=-Y \otimes X^{\otimes n}$ or other
possible forms. Besides, it is interesting to compare {\em
generalized Shor's codes using GHZ states} with {\em generalized
Shor's subsystem codes} \cite{bc06}, because they are constructed
under completely different motivations and terminologies.

 \subsection{Comments on channel-adapted codes and nonadditive codes}

  As has been shown in Subsection 2.2, GHZ states are maximally
  entangled multipartite quantum states and posses (simple and deep)
  algebraic, topological and geometric properties. These properties
  have been used to explore connections among quantum information and
   computation, the Yang--Baxter
  equation, low dimensional topology and differential geometry, see
  \cite{zrwwg07}. On the other hand,  GHZ states have been
  exploited to yield quantum error correction codes
  \cite{ssw07,lncy97, fletcher07,ls07}.  Some of them
  \cite{lncy97, fletcher07} can
  be described in the stabilizer formalism, i.e., additive codes,
  but many of them \cite{ssw07,ls07} are nonadditive codes which do
  not have classical analogues and do not satisfy the quantum Hamming
  bound (\ref{ham_bound}). Note that nonadditive codes using GHZ states
  are called {\em self-complementary codes} in  \cite{ssw07,ls07}.
  Some codes using GHZ states (either additive or nonadditive)
  are channel-adapted quantum error correction for
  the amplitude damping channel \cite{lncy97, fletcher07,ls07}.
   However, it remains an
  open problem how to create nonadditive codes using GHZ states
  in a unified theoretical framework, though interesting
  examples have been found. We expect that algebraic (or topological or
  geometrical) properties of GHZ states \cite{zrwwg07} are helpful
  for solving this problem. This is the motivation for our articulating
  quantum error correction codes using GHZ states.

  As examples, we analyze two codes \cite{lncy97, fletcher07} devised
  for the amplitude damping channel in our Hamiltonian formalism.
  They may be helpful for
  seeking a theoretical framework underlying nonadditive codes
  using GHZ states.  The first one \cite{lncy97} is the code
  $[4,1]$ having a logical qubit spanned by
  \eq
  |0\rangle_L =\frac 1 {\sqrt 2} (|0000\rangle)+|1111\rangle), \quad
  |1\rangle_L =\frac 1 {\sqrt 2} (|0011\rangle + |1100\rangle).
  \en
This  code can be rewritten as  the unitary basis transformation
$e^{-i\frac \pi 4 H}$,
 \eq
 (|0\rangle_L, |1\rangle_L)=e^{-i\frac \pi 4 H} (|0000\rangle, |0011\rangle)
 \en
where $iH=Y\otimes X^{\otimes 3}$ is called the almost-complex
structure and the Bell matrix $B=e^{-i\frac \pi 4 H}$ forms a
unitary braid representation, see \cite{zrwwg07}. Note that the Bell
matrix $B$ can generate all the GHZ states of four-qubit from the
product basis, i.e., it has all the information of GHZ states of
four-qubit. In this sense, the Hamiltonian formulation of the code
$[4,1]$ suggests: it appears to work only with two GHZ states of
four-qubit, but in fact it involves all other GHZ states of
 four-qubit through the Bell matrix $B$. This may be one of the reasons
why it violates the quantum Hamming bound (\ref{ham_bound}) but
works well for the amplitude-damping channel. The second example is
the code $[6,2]$ which has the form in the Hamiltonian formulation,
\eq
 (|00\rangle_L, |01\rangle_L, |10\rangle_L, |11\rangle_L) =
 e^{-i\frac \pi 4 H } (|000000\rangle, \quad |001001\rangle, |000110\rangle,
  |110000\rangle),
\en where $H=-iY\otimes X^{\otimes 5}$. A similar analysis can be
made: the unitary evolution operator $e^{-i\frac \pi 4 H }$ has the
information of all GHZ states of six-qubit, and hence the code
$[6,2]$ actually encodes two-qubit information into the entire
Hilbert space spanned by $2^6=64$ GHZ basis states instead of its
four-dimensional subspace. In this sense, we may understand why a
12-dimensional subspace can be encoded into the GHZ states of
8-qubit and protected by a nonadditive code \cite{ls07}.

Furthermore, we discuss how to protect a one-dimensional space by
quantum error correction codes using GHZ states. The quantum Hamming
bound condition (\ref{ham_bound}) at $t=0$ has a perfect solution in
terms of $B=1$, $N=3$ saturating the bound, i.e., $(1+1+2\times
3)\le 2^3$. Let us encode the one-dimensional space $|\psi\rangle$
using a GHZ state of three-qubit,   \eq
      |\psi\rangle_L =\frac 1 {\sqrt 2} (|000\rangle + |111\rangle)
      = e^{-i\frac \pi 4 H} |0\rangle^{\otimes 3}, \quad H =-i Y \otimes Y \otimes Y.
     \en
  There are three bit-flip errors, one phase-flip error and three bit-phase flip
  errors giving rise to the other seven GHZ states of three-qubit,
   \eq
 X_1 |\psi\rangle_L,   X_2 |\psi\rangle_L, X_3 |\psi\rangle_L, Z_1
 |\psi\rangle_L, Z_1 X_1 |\psi\rangle_L,   Z_2X_2 |\psi\rangle_L, Z_3X_3 |\psi\rangle_L,
   \en
 which form an orthonormal basis of ${\cal H}_2^{\otimes 3}$ together with
 $|\psi\rangle_L$. They can be recast in our Hamiltonian
 formulation
 \eqa
 && (Id, X_1, X_2, X_3, Z_1, Z_1 X_1, Z_2 X_2, Z_3 X_3)
 |\psi\rangle_L \nonumber\\
 &&=e^{-i\frac \pi 4 H}(|000\rangle,|011\rangle, |110\rangle, -|111\rangle, -|100\rangle,
       -|010\rangle, -|001\rangle )
\ena where four minus signs can be absorbed by rescaling the
 phase-flip $Z$-error. Besides, we can exploit the standard procedures of
encoding, error syndrome measurement, correcting and decoding
devised for the stabilizer codes \cite{gottesman97}.

 \section{Concluding remarks and outlook}

 Motivated by the problem how to construct a Hamiltonian model for
 nonadditive quantum error correction code, we suggest a new Hamiltonian
 formulation of quantum error correction code without appealing to
 its stabilizer formalism (if it exists).  As a remark, in Kitaev's
 fault tolerant schemes \cite{kitaev97}, thermal effect may not be easily
 surmountable \footnote{The author thanks Zohar Nussinov for the email
 correspondence on this point.} which can  destroy the expected
 fault-tolerance, see \cite{no07}, and hence seeking for new Hamiltonian
 models of quantum error correction codes remains a fundamental problem
 in the current research of quantum information and computation.

 In our Hamiltonian formulation, the unitary
 evolution operator at a specific time is a unitary basis
 transformation matrix from the product basis to the quantum error
 correction code, and we explain this by studying examples
 including Shor's nine-qubit code and its generalization. Remarkably,
 as this basis transformation matrix is a solution of the
 Yang--Baxter equation, the  quantum error correction
 code can be explained as a braiding operator, and the
 Hamiltonian model is an integrable model determined by the solution
 of the quantum Yang--Baxter equation. On the other
 hand, we articulate the topic called {\em quantum error correction codes
 using GHZ states}, in which new nonadditive codes and channel-adapted
 codes may be constructed with the help of beautiful properties of GHZ
 states.

 There still remain many interesting problems about our work in
 this paper. 1) Devise a theoretical
 framework for describing quantum error correction codes in the
 Hamiltonian formulation without involving the stabilizer formalism.
 Our typical examples are mainly based on algebraic properties of
 GHZ states. 2) Study physical properties of our Hamiltonian models
 and explore their applications to quantum error correction. As a
 comparison, quantum  error correction code in the Hamiltonian model
 based on the stabilizer formalism is its degenerate ground state, and
 there is a gap between this ground state and its first excited state.
 3) Study experimental realizations of these Hamiltonian models on optical
 lattices or atomic spin lattices \cite{pwskpw07,jbghhdlz07}. The
 unitary basis transformation from the product basis to GHZ
 states can be experimentally performed in view of the work
 \cite{klmt99}. 4) Study applications of these Hamiltonian
 models to develop a theoretical framework for {\em integrable quantum
 computation} or {\em topological-like quantum computation}. Note that the
 unitary evolution operator in the Hamiltonian model for a quantum error
 correction code using GHZ states can be chosen as a solution of
 the quantum Yang--Baxter equation. 5) About {\em quantum error
 correction codes using GHZ states}, there exist at least three types
 of questions to be answered: devise a general framework for
 yielding nonadditive codes; study quantum circuits for encoding,
 error correction and decoding; explore fault-tolerant quantum
 computation with the help of properties of GHZ states, for example,
 channel-adapted fault-tolerant quantum computation for the amplitude
 damping channel.

  \section*{Acknowledgements}

   The author thanks Yong-Shi Wu for support,
   thanks Peter W. Shor for helpful comment, and he is in part
   supported by the seed funding of University of Utah and NSF-China
   Grant-10605035.

   He was a participant in First International Conference on
   Quantum Error Correction (University of Southern California,
   Los Angeles, USA, Dec. 17-21, 2007), and thanks Dave Bacon and
   John Preskill for a relevant discussion.

\end{document}